\documentclass[10pt,aps,prl,twocolumn,superscriptaddress,showpacs]{revtex4-1}

\usepackage[ansinew]{inputenc}
\usepackage{scrextend} 
\usepackage{epsfig}
\usepackage{amsmath}
\usepackage{graphicx}
\usepackage{subfigure}
\usepackage{color}
\usepackage{times}
\usepackage{bm}
\usepackage{multirow}
\usepackage{hyperref}
\usepackage{verbatim}
\usepackage{siunitx}

\begin{document}

\title{X-ray resonant photoexcitation: line widths and energies of K\texorpdfstring{$_{\alpha}$}{\_alpha} transitions in highly charged Fe ions}
%
\author{J.~K.~Rudolph}\email[]{jan.rudolph@physik.uni-giessen.de}
\affiliation{Max-Planck-Institut f\"ur Kernphysik, Saupfercheckweg 1, 69117 Heidelberg, Germany}
\affiliation{Institut f\"ur Atom- und Molek\"ulphysik, Justus-Liebig-Universit\"at Gie{\ss}en, Leihgesterner Weg 217, 35392 Gie{\ss}en, Germany}

\author{S.~Bernitt}\affiliation{Max-Planck-Institut f\"ur Kernphysik, Saupfercheckweg 1, 69117 Heidelberg, Germany}
\author{S.~W.~Epp}\affiliation{Max Planck Advanced Study Group, CFEL, Notkestra{\ss}e 85 , 22607 Hamburg, Germany}
\author{R.~Steinbr\"ugge}\affiliation{Max-Planck-Institut f\"ur Kernphysik, Saupfercheckweg 1, 69117 Heidelberg, Germany}
\author{C.~Beilmann}\affiliation{Max-Planck-Institut f\"ur Kernphysik, Saupfercheckweg 1, 69117 Heidelberg, Germany}
\affiliation{Physikalisches Institut, Ruprecht-Karls-Universit\"at Heidelberg, Im Neunheimer Feld 226, 69120 Heidelberg, Germany}
\author{G.~V.~Brown}\affiliation{Lawrence Livermore National Laboratory, 7000 East Avenue, Livermore, CA 94550, USA}
\author{S.~Eberle}\affiliation{Max-Planck-Institut f\"ur Kernphysik, Saupfercheckweg 1, 69117 Heidelberg, Germany}
\author{A.~Graf}\affiliation{Lawrence Livermore National Laboratory, 7000 East Avenue, Livermore, CA 94550, USA}
\author{Z.~Harman}\affiliation{Max-Planck-Institut f\"ur Kernphysik, Saupfercheckweg 1, 69117 Heidelberg, Germany}
\affiliation{ExtreMe Matter Institute (EMMI), Planckstra{\ss}e 1, 64291 Darmstadt, Germany}
\author{N.~Hell}\affiliation{Lawrence Livermore National Laboratory, 7000 East Avenue, Livermore, CA 94550, USA}
\affiliation{Dr. Karl Remeis-Observatory \& ECAP, Universit\"at Erlangen N\"urnberg, Sternwartstra{\ss}e 7, 96049 Bamberg, Germany }
\author{M.~Leutenegger}\affiliation{NASA/Goddard Space Flight Center, 8800 Greenbelt Road, Greenbelt, MD 20771, USA}
\affiliation{Center for Space Sciences \& Technology, University of Maryland, Baltimore County, Baltimore, MD 21250, USA}
\author{A.~M\"uller}\affiliation{Institut f\"ur Atom- und Molek\"ulphysik, Justus-Liebig-Universit\"at Gie{\ss}en, Leihgesterner Weg 217, 35392 Gie{\ss}en, Germany}
\author{K.~Schlage}\affiliation{Deutsches Elektronen-Synchrotron (PETRA III), Notkestra{\ss}e 85, 22607 Hamburg, Germany}
\author{H.-C.~Wille}\affiliation{Deutsches Elektronen-Synchrotron (PETRA III), Notkestra{\ss}e 85, 22607 Hamburg, Germany}
\author{H.~Yavas}\affiliation{Deutsches Elektronen-Synchrotron (PETRA III), Notkestra{\ss}e 85, 22607 Hamburg, Germany}
\author{J.~Ullrich}\altaffiliation{present address: Physikalisch Technische Bundesanstalt (PTB), Bundesallee 100, 38116 Braunschweig, Germany}\affiliation{Max-Planck-Institut f\"ur Kernphysik, Saupfercheckweg 1, 69117 Heidelberg, Germany}
\author{J.~R.~{Crespo L\'opez-Urrutia}}\affiliation{Max-Planck-Institut f\"ur Kernphysik, Saupfercheckweg 1, 69117 Heidelberg, Germany}


\date{\today}
\begin{abstract}
Photoabsorption by and fluorescence of the K$_{\alpha}$ transitions in highly charged iron ions are essential mechanisms for X-ray radiation transfer in astrophysical environments. We study photoabsorption due to the main K$_{\alpha}$ transitions in highly charged iron ions from heliumlike to fluorinelike (Fe$^{24+...17+}$) using monochromatic X-rays around \SI{6.6}{\kilo\electronvolt} at the PETRA III synchrotron photon source. Natural linewidths were determined with hitherto unattained accuracy. The observed transitions are of particular interest for the understanding of photoexcited plasmas found in X-ray binaries and active galactic nuclei.
\end{abstract}
 

\maketitle
The spectrum of highly charged iron ions provides rich information on the dynamics of X-ray binaries
\cite{Ezuka1999,Matt1996,Martocchia1996,Nandra2006,Reynolds2008,Ebisawa1996,Fabian1995,Yaqoob2001,Page2004,Lee2002,Miller2002}. 
Owing to their high transition rate, low intergalactic absorption, and the high relative abundance, the Fe K$_{\alpha}$ spectral features have been detected in a variety of celestial sources and are of particular interest to the study of high mass X-ray binaries and active galactic nuclei and their accretion disks. The K$_{\alpha}$ features often provide the last recognizable spectral signature of the accreted material.  They are generated by the reflection or absorption of X-rays emitted from the vicinity of the compact object or central black hole and, potentially, from the relativistic jets found in AGN and micro-quasars \cite{Garcia2013,Miller2008a}. Hence, not only are properties of the accretion disk encoded in the Fe K$_{\alpha}$ features, so too are physical properties of the black hole or neutron star. For example, measured Doppler shifts produced by the rotating accretion disk, the relativistic beaming, gravitational bending, and even subtler effects from general relativity, such as those arising from the black hole's spin, all leave imprints on the spectral shape of the Fe K$_{\alpha}$ transitions \cite{Eracleous1996,Lubinski2001,Branduardi2001,Popovic2003,Nandra2007,Corral2008}. A full, quantitative understanding would require not only a consistent modeling of these effects, but also an accurate and complete set of, in the case of photoionized sources, laboratory-tested oscillator strengths and transition energies on which the diagnostics are based.

Experimentally, the Fe K$_{\alpha}$ transitions have been studied as emission lines from thermal plasmas, in tokamaks \cite{Bitter1979,Briand1982,Beiersdorfer1993} and by means of Lawrence Livermore National Laboratory's EBIT-II electron beam ion trap \cite{decaux1997a,decaux1995a}, using a mono-energetic electron beam. 
High resolution K$_{\alpha}$ line spectra of highly charged Ar ions were also obtained with an electron cyclotron resonance ion source \cite{Amaro2012, Szabo2013}.
In all cases, the excitation of the $n=2$ upper level was produced by electron impact excitation, or more complex processes such as inner shell ionization, or dielectronic recombination. 

\begin{figure}[tb!]
 \centering
	\includegraphics[width = 0.85\columnwidth]{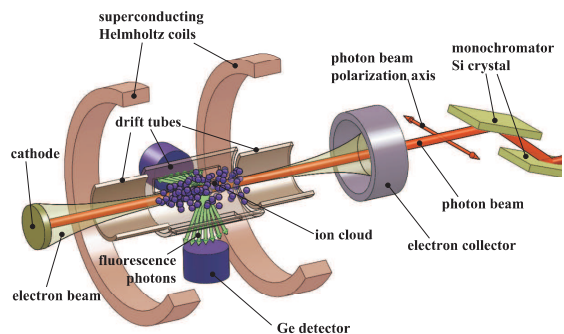}
	\caption{
		Scheme of the experimental setup . An electron beam accelerated toward the trap center and compressed by a coaxial \SI{6}{\tesla} magnetic field produces Fe$^{24+}$ ions which are then irradiated by a monochromatic X-ray beam. Resonant excitation leads to fluorescence emission, which is registered by germanium photon detectors.
		\label{fig:setup} }
\end{figure}

\begin{figure}[b!]
\centering
	\includegraphics[width = \columnwidth]{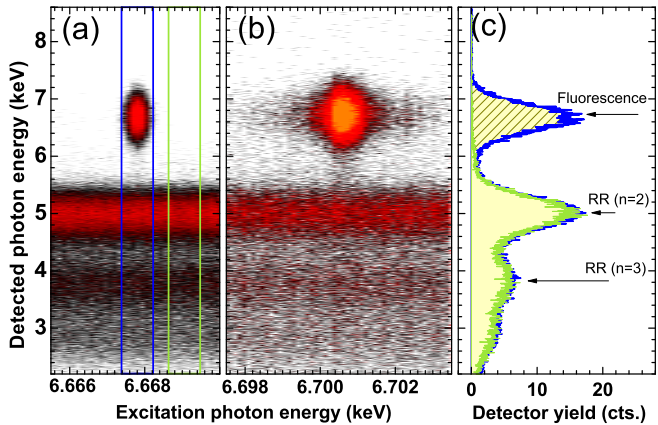}
	\caption{
		\textcolor{blue}{(color online)} Typical emission spectra. (a),(b): Detector yield plotted against monochromator energy. Strong signals due to energy dependent fluorescence photons, well separated from energy independent background at lower energies are visible. On top of the constant background caused by electron impact excitation there is an enhancement at \SI{5}{\kilo\electronvolt} and \SI{3.9}{\kilo\electronvolt}. This is due to radiative recombination (RR) by the electron beam into the $n=2$ and $n=3$ shell. Below \SI{3}{\kilo\electronvolt}, strong absorption by the beryllium windows installed in front of the detectors suppresses the signal. (c): The blue curve shows the detector yield in the resonance regime of spectrum (a) and the green curve in a region off-resonance of the same width.}
\label{fig:EIE_bg}
\end{figure}

In the present experiment a high-fluence X-ray photon beam is used for excitation, in order to isolate and probe the fundamental structure of highly charged iron ions in the absence of any collisional excitation mechanism. We employ the monoenergetic photon beam produced at the PETRA III synchrotron at DESY to excite K$_\alpha$ transitions in Fe ions, produced and trapped by MPIK's FLASH-EBIT (see FIG. \ref{fig:setup}). FLASH-EBIT, described in detail by Epp et al. \cite{Epp2010}, was designed for photonic studies of highly charged ions (HCI) and has been used in several experiments at other synchrotrons and free-electron lasers \cite{Epp2007, Simon2010, Simon2010b,Bernitt2012}. At its most basic, FLASH-EBIT uses a focused electron beam of several hundred milliamperes to ionize, and trap the ions. The electron beam is compressed by means of a magnetic field of \SI{6}{\tesla} at the trap center to a radius below \SI{50}{\micro\meter}. An electrostatic potential well formed at the center of three drift-tube electrodes provides longitudinal trapping of the ions. 
In the cylindrical trapping region, an ion cloud of \SI{50}{\milli\meter} length and $\sim$\SI{200}{\micro\meter} diameter is stored at an areal density of up to \SI{e11}{ions\per\centi\square\meter}. The maximum charge state achievable is predominantly determined by the electron beam accelerating potential, and narrow charge state distributions are routinely achieved. For the experiments presented here, neutral iron atoms are brought into the trapping region by means of a molecular beam of iron pentacarbonyl, which undergoes dissociation and ionization at the crossing point with the electron beam. Once the ions are generated, they are radially trapped by the negative space charge of the beam and axially by the drift tubes. By limiting the electron beam energy to \SI{3.24}{\kilo\electronvolt}, the highest charge state produced was Fe$^{24+}$. At this relatively low electron beam energy, direct electron impact excitation of the $n = 1$ to $2$ transitions in the Fe ions was precluded. Photorecombination-induced background was also absent in the Fe K$_{\alpha}$ energy region, since the sum of the ionization potential of $n = 2$ shell vacancies and the kinetic energy of the electron amounts to less than \SI{5}{\kilo\electronvolt}. Typical emission spectra with fluorescence photons separated from photorecombination background are shown in FIG. \ref{fig:EIE_bg}.

Trapped ions were resonantly excited using the high-fluence photon beam at the Dynamics Beamline P01 at the PETRA III synchrotron X-ray source. 
At P01, the ultra intense X-ray beams are generated using two \SI{5}{m} long undulators with a period length of \SI{32}{\milli\meter}.  
Because of the magnetic field orientation of the undulators, a horizontally polarized photon beam is produced. The beamline is dedicated, in general, to inelastic X-ray and nuclear resonant scattering; both methods require high flux.
A double-crystal, high heat load monochromator (HHLM) equipped with two Si crystal pairs is employed. 
The Si(111) cut crystal yields a resolution of about \SI{1}{\electronvolt} full-width-at-half-maximum (FWHM) at  \SI{6.6}{\kilo\electronvolt}, and with the Si(311) cut \SI{0.2}{\electronvolt} is achieved,  albeit with reduced beam intensity. 
After the HHLM the flux was up to \num{e12} photons per second on a spot of less than \SI{1}{\milli\meter} diameter and a resolving power of $E/{\Delta E} \geq 20000$ was achieved. 

\begin{table*}[ht!]
     \centering
     \renewcommand{\arraystretch}{1.25}
	\begin{tabular}{p{1.0cm}p{0.7cm}p{2.5cm}p{2.4cm}p{2.8cm}p{2.1cm}p{0.7cm}p{1.05cm}p{0.7cm}p{1.9cm}p{0.6cm}}
	\hline\hline
		  Ion&Line &Initial state &Final state &Experiment &Theory&&Theory&&Exp.\\\hline
		Fe$^{24+}$& $w$&  $1s^2\,^1S_{0}$&$ 1s\ 2p\,^1P_{1}$	&$6700.549\,(5)\,(70)$ 	&$6700.4347\,(11)$&\cite{Artemyev2005}&\num{6700.4} &\cite{Palmeri2003}&$6700.8$&\cite{Beiersdorfer1993} \\
		   & &	&& &$6700.490$&\cite{Plante1994}&$6700.4$&\cite{Johnson1995}&$6700.4$&\cite{decaux1997a}\\
		   & &	&& &&&&&$6700.9$&\cite{Briand1984}\\
		 Fe$^{24+}$& $y$& $1s^2\,^1S_{0}$&$ 1s\ 2p\,^3P_{1}$	&$6667.671\,(3)\,(69)$ 	&$6667.5786\,(12)$&\cite{Artemyev2005}&\num{6667.6} &\cite{Palmeri2003}&$6667.9\,(4)$&\cite{Beiersdorfer1993}\\
		   &&	 &	&	&$6667.629$&\cite{Plante1994}&$6667.6$&\cite{Johnson1995}&$6667.5$&\cite{decaux1997a}\\
		   		   &&	 &	&	&&&&&$6667.5$&\cite{Briand1984}\\
		Fe$^{23+}$& $t$&  $1s^2\,2s\,^2S_{1/2}$&$ 1s\ 2s\ 2p\,^2P_{1/2}$	&$6676.202\,(3)\,(69)$ 	&$6676.129\,(47)$&\cite{Yerokhin2012}&\num{6676.4} &\cite{Palmeri2003}&$\langle6676.8\,(7)\rangle$ &\cite{Beiersdorfer1993}\\
		   & & &	&	&$6675.8 $&\cite{Safronova2010}&&&$6676.3$&\cite{decaux1997a}\\
		Fe$^{23+}$& $q$& $1s^2\,2s\,^2S_{1/2}$&$ 1s\ 2s\ 2p\,^2P_{3/2}$	&$6662.240\,(6)\,(69) $	&$6662.188\,(11) $&\cite{Yerokhin2012}&\num{6661.9} &\cite{Palmeri2003}&$6662.1\,(5)$	&\cite{Beiersdorfer1993}\\
		   & &	&	&							&$6661.9 $	&\cite{Safronova2010}&&&$6662.2$&\cite{decaux1997a}\\
		 Fe$^{23+}$& $r$& $1s^2\,2s\,^2S_{1/2}$&$ 1s\ 2s\ 2p\,^2P_{1/2}$	&$6652.826\,(3)\,(69) 	$&$6652.776\,(25)$ &\cite{Yerokhin2012}&\num{6653.5}&\cite{Palmeri2003}&$\langle6654.2\,(7)\rangle$	&\cite{Beiersdorfer1993} \\
		   &  	&	&	&						&$6652.6 	$&\cite{Safronova2010}&&&$6652.5$&\cite{decaux1997a}\\
		Fe$^{23+}$& $u$& $1s^2\,2s\,^2S_{1/2}$&$ 1s\ 2s\ 2p\,^4P_{3/2}$&$6616.629\,(4)\,(68) $&$6616.559\,(11)$&\cite{Yerokhin2012}&\num{6616.7} &\cite{Palmeri2003}&$\langle6617.9\,(1.2)\rangle$&\cite{Beiersdorfer1993}\\
		&&&&&&&&&$6616.6$&\cite{decaux1997a}\\
		Fe$^{22+}$& $E1$& $1s^2\,2s^2\,^1S_{0}$&$ 1s\ 2s^2\ 2p\,^1P_{1}$	& $6628.804\,(5)\,(68)$ 	&$6631.057$&\cite{Shlyaptseva1998}&\num{6628.7} &\cite{Palmeri2003}	&$6628.9\,(3)$&\cite{Beiersdorfer1993}\\
						&  	&	&&&	$6627.4\footnotemark[1]\ /\ 6628.3$\footnotemark[2]&&$6627.39 	$& \cite{Chen1985}&$6628.7$&\cite{decaux1997a}\\
		Fe$^{22+}$& $E2$& $1s^2\,2s^2\,^1S_{0}$&$ 1s\ 2s^2\ 2p\,^3P_{1}$	& $6597.858\,(3)\,(67)$&$6596.55 	$&\cite{Chen1985}&\num{6595.8} &\cite{Palmeri2003}\\
				&  	&	&	&&$6596.1\footnotemark[1]\ /\ 6597.7 	$\footnotemark[2]&\\
		\multirow{2}{*}{Fe$^{21+}$}& \multirow{2}{*}{$B$}& \multirow{2}{*}{$1s^2\,2s^2\ 2p\,^2P_{1/2}$}&$ 1s\ 2s^2\ 2p^2\,^2P_{1/2}$&$\langle6586.085\,(3)\,(67)\rangle$ &$6586.3\footnotemark[1]\ /\ 6585.1$\footnotemark[2]&&\num{6586.3} &\cite{Palmeri2003} &$\langle6585.9\,(5)\rangle$&\cite{Beiersdorfer1993}\\
		&&&$ 1s\ 2s^2\ 2p^2\,^2D_{3/2}$&$\langle6586.085\,(3)\,(67)\rangle$&$6587.0\footnotemark[1]\ /\ 6585.8$\footnotemark[2]&&\num{6586.5} &\cite{Palmeri2003}&$\langle6585.7\rangle$&\cite{decaux1997a}\\
		&&&&&$\langle6587.2\rangle$& \cite{Shlyaptseva1996}\\
		Fe$^{20+}$& $C1$& $1s^2\,2s^2\ 2p^2\,^3P_{0}$&$ 1s\ 2s^2\ 2p^3\,^3D_{1}$&$6544.225\,(4)\,(66)$&$6544.8\footnotemark[1]\ /\ 6544.0$\footnotemark[2]&&\num{6543.6} &\cite{Palmeri2003}&$\langle6544.6\,(9)\rangle$&\cite{Beiersdorfer1993}\\
		&&&&&&&&&$\langle6544.4\rangle$&\cite{decaux1997a}\\
		Fe$^{20+}$& $C2$& $1s^2\,2s^2\ 2p^2\,^3P_{0}$&$ 1s\ 2s^2\ 2p^3\,^3S_{1}$&$6556.879\,(16)\,(66)$&$6557.3\footnotemark[1]\ /\ 6556.3$\footnotemark[2]&&\num{6555.0} &\cite{Palmeri2003}\\
		Fe$^{19+}$& $N1$& $1s^2\,2s^2\ 2p^3\,^4S_{3/2}$&$ 1s\ 2s^2\ 2p^4\,^4P_{5/2}$&$6497.067\,(5)\,(65)$&$6497.5\footnotemark[1]\ /\ 6497.2$\footnotemark[2]&&\num{6496.6} &\cite{Palmeri2003}&$\langle6497.7\,(1.4)\rangle$&\cite{Beiersdorfer1993}\\
		&&&&&&&&&$6497.3$&\cite{decaux1997a}\\
		Fe$^{19+}$& $N2$& $1s^2\,2s^2\ 2p^3\,^4S_{3/2}$&$ 1s\ 2s^2\ 2p^4\,^2P_{3/2}$&$6506.845\,(7)\,(65)$&$6507.3\footnotemark[1]\ /\ 6506.9$\footnotemark[2]&&\num{6506.0} &\cite{Palmeri2003}&$\langle6509.6\,(1.4)\rangle$&\cite{Beiersdorfer1993}\\
		&&&&&&&&&$\langle6509.1\rangle$&\cite{decaux1997a}\\
		Fe$^{19+}$& $N3$& $1s^2\,2s^2\ 2p^3\,^4S_{3/2}$&$ 1s\ 2s^2\ 2p^4\,^4P_{1/2}$&$6509.133\,(14)\,(65)$&$6509.6\footnotemark[1]\ /\ 6509.1$\footnotemark[2]&&\num{6508.1} &\cite{Palmeri2003}&$\langle6509.6\,(1.4)\rangle$&\cite{Beiersdorfer1993}\\
		&&&&&&&&&$\langle6509.1\rangle$&\cite{decaux1997a}\\
		Fe$^{18+}$& $O1$& $1s^2\,2s^2\ 2p^4\,^3P_{2}$&$ 1s\ 2s^2\ 2p^5\,^3P_{2}$&$6466.900\,(14)\,(64)$&$6467.4\footnotemark[1]\ /\ 6466.5$\footnotemark[2]&&\num{6564.4} &\cite{Palmeri2003}&$6467.6\,(1.7)$&\cite{Beiersdorfer1993}\\
		&  	&	&	&						&$6466.3 	$&\cite{Shlyaptseva1996}&&&$\langle6466.5\rangle$&\cite{decaux1997a}\\
		Fe$^{18+}$& $O2$& $1s^2\,2s^2\ 2p^4\,^3P_{2}$&$ 1s\ 2s^2\ 2p^5\,^3P_{1}$&$6474.318\,(33)\,(64)$&$6474.9\footnotemark[1]\ /\ 6473.9$\footnotemark[2]&&\num{6473.0} &\cite{Palmeri2003}&$\langle6472.7\,(2.7)\rangle$&\cite{Beiersdorfer1993}\\
		  &  	&	&	&						&$6473.7 	$&\cite{Shlyaptseva1996}&&&$\langle6474.7\rangle$&\cite{decaux1997a}\\
		Fe$^{17+}$& $F$& $1s^2\,2s^2\ 2p^5\,^2P_{3/2}$&$ 1s\ 2s^2\ 2p^6\,^2S_{1/2}$&$6435.239\,(14)\,(63)$&$6435.7\footnotemark[1]\ /\ 6434.6$\footnotemark[2]&&\num{6434.8} &\cite{Palmeri2003}&$6436.1\,(2.0)$&\cite{Beiersdorfer1993}\\
				&&&&&&&&&$6434.8$&\cite{decaux1997a}\\\hline\hline
		\footnotetext[1]{\mbox{Our theoretical results obtained in the framework of the multiconfiguration Dirac-Fock (MCDF) method \cite{Simon2010}}}
		\footnotetext[2]{\mbox{Our theoretical results using Flexible Atomic Code (FAC) with the standard configuration-interaction package \cite{Gu2008}.}}
\end{tabular}
     \caption{X-ray transitions of heliumlike to fluorinelike iron ions resonantly excited from the ground state with synchrotron radiation. X-ray fluorescence was detected as a function of photon energy. Energies are given in units of \si{\electronvolt}. The calibration is based on the absorption edge technique. The experimental uncertainties are shown as (statistical)(systematic). Relative energies are not affected by the systematic uncertainty which accounts for a shift of the absolute scale. Angle brackets enclose results affected in their accuracy by line blends.}
\label{tab:FeWY-Lines}
\end{table*}

\begin{figure}[!tb]
\begin{center}
\includegraphics[width = \columnwidth]{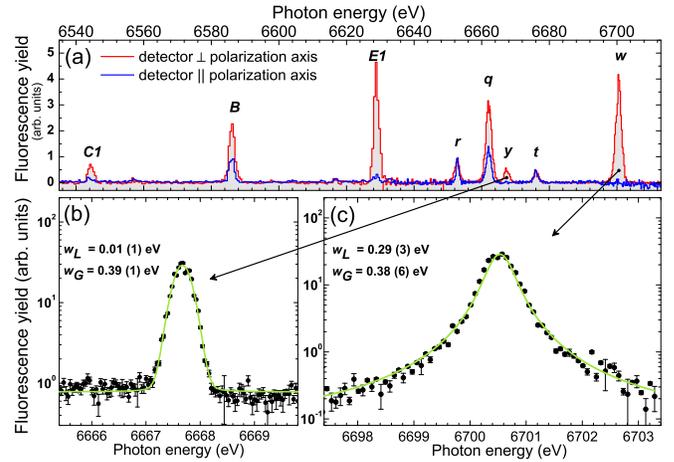}
\caption{\label{fig:FeWY}
(a): Overview of the resonance fluorescence spectrum of heliumlike, lithiumlike, berylliumlike, boronlike and carbonlike Fe ions using two X-ray photon detectors. Alignment of detectors with respect to the polarization of the incoming photon beam: (red) perpendicular and (blue) parallel. (b): High-resolution scan over the intercombination line ($y$)\,$1s^2\,^1S_{0}\rightarrow 1s\ 2p\,^3P_{1}$. (c): High-resolution scan over the resonance transition ($w$)\,$1s^2\,^1S_{0}\rightarrow 1s\ 2p\,^1P_{1}$. The fitted Gaussian and the Lorentzian widths $w_G$ and 
$w_L$ from a Voigt profile of the transitions are shown.}
\end{center}
\end{figure}

The photon beam at P01 was brought into FLASH-EBIT through a beryllium window installed in the electron beam collector chamber. Following our earlier work at advanced light sources, the P01 photon beam is superimposed along the EBIT axis  with the trapped ion cloud (FIG. \ref{fig:setup}). Overlap of the photon beam with the ion cloud was established by use of a retractable scintillator inserted into the trap region and viewed by an image-intensified camera. Both the photon beam and the electron beam fluoresce the scintillator, and maximum overlap is achieved by moving the entire FLASH-EBIT with remote controlled positioners. 

Fluorescent photons are detected using two liquid nitrogen cooled germanium solid-state detectors (\SI{1000}{\square\milli\meter} area each) with an intrinsic resolution of approximately \SI{500}{\electronvolt} FWHM at \SI{6}{\kilo\electronvolt}. One detector was aligned parallel and one perpendicular to the X-ray beam polarization axis and both perpendicular to the propagation direction. Beryllium windows mounted radially provided a detection solid angle of $\Omega \approx \SI{1e-3}{\steradian}$ for both detectors.  The energy resolution of these detectors allows to separate the fluorescence signal from the lower photon energy X-ray background generated by electron beam excitation of the trapped ions and by radiative recombination processes.

For these measurements, X-ray fluorescence was measured using the HHLM to scan the X-ray photon beam across the energy band containing the K$_{\alpha}$ transitions. During the first scans, the energy band between \SIrange{6530}{6710}{\electronvolt} was covered using the Si(111) reflection to maximize beam intensity. The observed spectra for both the parallel and perpendicular detectors are shown in FIG. \ref{fig:FeWY}. The spectra include, for heliumlike iron, the $w$ and $y$ transitions, for the lithiumlike the $q$, $t$ and $r$ transitions, and also further resonant transitions in berylliumlike, boronlike, and carbonlike iron ions (TAB.~\ref{tab:FeWY-Lines}). Labels for heliumlike and lithiumlike transitions are used as in \cite{Gabriel1972}. Because the incident photon beam is polarized, the resulting fluorescence spectra are emitted non-isotropically depending on the transition type. Specifically, $\Delta J=1$ transitions have a radiation pattern that peaks at \SI{90}{\degree} to the polarization axis, while radiation from $\Delta J=0$ transitions is isotropic.

To determine transition energies we calibrated the energy scale of the monochromator by measuring the absorption edge spectra of metal foils. The intensity of the photon beam after it passed through each foil was measured using a gas proportional chamber, and represented as a function of the photon beam energy (Bragg angle of the crystal). Digital values of the photon energy selected by the monochromator are provided by a data server in real time. The transmitted photon intensity is normalized to the signal taken before the absorption foil, also using a gas proportional counter. A corrected lattice spacing is used which is more accurate than the control software value by taking into account the crystals thermal expansion due to the liquid nitrogen cooling \cite{Lyon1977}. 
Edges are identified by calculating the photon energy derivative of the transmitted intensity signal. 
As shown in \cite{Deslattes2003}, indirectly determined K-edge energies for Mn and Fe have seven and twenty times larger uncertainties, respectively, than the direct wavelength measurements of Kraft et al. ($\pm$\SI{20}{\milli\electronvolt})  \cite{Kraft1996}. 
Theoretical values given in \cite{Deslattes2003} disagree by \SI{15}{\electronvolt} from the experiment due to unknown solid state effects. Therefore, we use the K-edges of manganese, iron, cobalt, nickel and copper by Kraft et al. as references.
This results in an angular offset for which one has to correct the Bragg angles provided by the PETRA III data acquisition system since the monochromator has no absolute zero position. In Tab. \ref{tab:FeWY-Lines} we compare our results with several theoretical calculations, which especially for heliumlike and lithiumlike iron, agree within the experimental uncertainty with our measurement.

Some of the lines were then investigated at higher resolution in further scans using the monochromator with the Si(311) cut crystal, thereby improving the photon beam resolution by a factor of \num{5}. Additionally, the effect of thermal Doppler line broadening was reduced by cooling the trapped ions evaporatively \cite{Levine1989}. We decreased the thermal broadening to be comparable to the natural line widths of the permitted $1s$-$2p$ transitions while still producing sufficient fluorescence signal. High resolution scans of $w$ and $y$ are shown in the bottom panels of FIG. \ref{fig:FeWY}.
\begin{table}[thb!]
	\centering
		\begin{tabular}{p{0.1\columnwidth}p{0.2\columnwidth}p{0.22\columnwidth}p{0.1\columnwidth}p{0.11\columnwidth}p{0.10\columnwidth}}
			\hline\hline
			\multirow{2}{*}{Line} & \multirow{2}{*}{Experiment}   & \multirow{2}{*}{Theory}  & & MCDF \\
			&&&rad.&Auger&total\\\hline
			$w$ & $311\,(10)$ & $301$ \cite{Drake1979}&$301$ &\quad$0$&\ \ $301$\\
								&&$301$ \cite{Johnson1995}\\
									&& $303$ \cite{Palmeri2003} \\	
									&& $315$ \cite{Hata1984}\\
			$q$ & $255\,(31)$ & $310$ \cite{Palmeri2003}	&$312$&$0.081$&\ \  $312$\\
			$r$ & $250\,(11)$ &   $226$ \cite{Palmeri2003} & $208$&$\ \ 32$&\ \ $241$ \\
			$t$ & $131\,(29)$ &   $167$ \cite{Palmeri2003} & $112$&$\ \ 52$&\ \ $163$\\
			$E1$ & $437\,(12)$ &   $382$ \cite{Palmeri2003} &$292$&$\ \ 95$&\ \ $387$\\
			$E2$ & 	$178\,(34)$		&  $149$ \cite{Palmeri2003}& \ \ $31$& $136$&\ \ $167$\\
			$C1$ & $524\,(12)$ &   $499$ \cite{Palmeri2003} &$218$&$334$ &\ \ $552$\\
			$C2$ & $385\,(207)$ &   $611$ \cite{Palmeri2003} &$412$& $135$&\ \ $547$ \\
			$N1$ & $565\,(45)$ &   $498$ \cite{Palmeri2003}  &$146$& $400$&\ \ $546$\\
			$N2$ & $594\,(50)$ &   $504$ \cite{Palmeri2003}  &$164$& $338$&\ \ $502$\\
			$N3$ & $570\,(109)$ &   $505$ \cite{Palmeri2003}  &$155$& $360$ &\ \ $515$\\
			$O1$ & $859\,(229)$ &   $756$ \cite{Palmeri2003}  &$264$& $529$&\ \ $793$\\
			$O2$ & $772\,(228)$ &   $785$ \cite{Palmeri2003}  &$304$& $518$&\ \ $822$\\
			$F$ & $998\,(203)$ &   $989$ \cite{Palmeri2003}  &$351$& $651$&$\,1020$\\	\hline\hline
		\end{tabular}
	\caption{Experimentally determined natural line widths $w_\text{L}$ of several $1s\rightarrow 2p$ transitions in comparison with predicted values, in units of \si{\milli\electronvolt}. Uncertainty given as \SI{1}{$\sigma$}. For line $B$ we can not determine a natural line width as there is a blend of two transitions. The line width of the intercombination lines $y$ and $u$ is dominated by Doppler broadening which also prevents a natural line width determination. The estimated uncertainty of MCDF widths is $\approx$\SI{10}{\percent}.}
	\label{tab:linewidth}
\end{table}

Natural line widths were determined by fitting the line shapes with a Voigt profile, as done by Beiersdorfer et al. \cite{Beiersdorfer1996-2} for Cs$^{45+}$ under electron impact excitation. A weighted average of the Lorentzian widths $w_\text{L}$ in all energy scanned spectra is taken. The intercombination transition $^3P_{1}$ $\rightarrow$ $^1S_{0}$ in heliumlike Fe forming line \textit{y} has a negligibly small natural line width in comparison to the Doppler width of this line. We compare the Doppler widths (Gaussian widths of the Voigt profiles) of the heliumlike line $w$ with the line widths of $y$ for spectra taken at the same EBIT conditions and find perfect agreement within the error bars, which supports our results for the natural line widths. Our natural line width measurements are summarized in TAB.~\ref{tab:linewidth} which also displays our theoretical results obtained in the framework of the MCDF method \cite{Simon2010}. The good agreement with theory extends even to the more complex ions, for which one might have expected larger deviations.

The strong K$_{\alpha}$ transitions of the highly charged iron ions are essential for astrophysical plasma diagnostics. With the present measurements, their resonant photoabsorption has been directly measured, and state-of-the-art predictions for both their transition probabilities and energies have been benchmarked against, and confirmed by, them with satisfactory results. As for the transition energies, our systematic uncertainty enlarges the total final error bar to a typical level of \SI{70}{\milli\electronvolt} while the energy differences show ten times smaller error bars. These results will be particularly useful when interpreting astrophysical X-ray spectra, especially from active galactic nuclei and high mass X-ray binaries, where absorption and reflection of X-rays from accretion disk is directly related to the physical properties of potentially every physical property of the system. The data will be especially useful for benchmarking theoretical calculations of atomic structure and spectral models used to interpret spectra from celestial sources, such as \emph{XSTAR} \cite{Bautista2001} and \emph{Cloudy} and are timely given the launch of the X-ray Calorimeter Spectrometer \cite{mitsuda2010a} on \emph{ASTRO-H} \cite{takahashi2012a} in 2015.

\begin{acknowledgments}
The research leading to these results was supported by Deutsche Forschungsgemeinschaft (DFG). J.K.R., N.H. and C.B. received funding from DESY, a member of the Helmholtz Association (HGF). 
Work by LLNL was performed under the auspices of the U.S. Department of Energy under Contract No. DE-AC52-07NA27344 and supported by NASA grants.
N.H. also recieved funding from Bundesministerium für Wirtschaft und Technologie (BMWi) under Deutsches Zentrum für Luft- und Raumfahrt (DLR) Grant 50\,OR\,1113. 
\end{acknowledgments}


\bibliography{PetraIII2012FeKalphaB}

\end{document}